\documentclass[aps,prb,twocolumn,showpacs,preprintnumbers,amsmath,amssymb,superscriptaddress]{revtex4}%

\usepackage{amsmath, amsthm, amssymb}
\usepackage{graphicx}
\usepackage{dcolumn}
\usepackage{bm}
\usepackage{color}

\begin{document}

\preprint{PREPRINT (\today)}

\newpage

\title{Coexistence of Superconductivity and Magnetism in FeSe$_{1-x}$ under Pressure}

\author{M.~Bendele}
\email{markus.bendele@physik.uzh.ch}
\affiliation{Physik-Institut der Universit\"{a}t Z\"{u}rich, Winterthurerstrasse 190, CH-8057 Z\"{u}rich, Switzerland}
\affiliation{Laboratory for Muon Spin Spectroscopy, Paul Scherrer Institut, CH-5232 Villigen PSI, Switzerland}

\author{A.~Ichsanow}
\affiliation{Physik-Institut der Universit\"{a}t Z\"{u}rich, Winterthurerstrasse 190, CH-8057 Z\"{u}rich, Switzerland}

\author{Yu.~Pashkevich}
\affiliation{A. A. Galkin Donetsk Phystech NASU, 83114 Donetsk, Ukraine}

\author{L.~Keller}
\affiliation{Laboratory for Neutron Scattering, Paul Scherrer Institut, CH-5232 Villigen PSI, Switzerland}

\author{Th.~Str\"assle}
\affiliation{Laboratory for Neutron Scattering, Paul Scherrer Institut, CH-5232 Villigen PSI, Switzerland}

\author{A.~Gusev}
\affiliation{A. A. Galkin Donetsk Phystech NASU, 83114 Donetsk, Ukraine}

\author{E.~Pomjakushina}
\affiliation{Laboratory for Developments and Methods, Paul Scherrer Institut, CH-5232 Villigen PSI, Switzerland}

\author{K.~Conder}
\affiliation{Laboratory for Developments and Methods, Paul Scherrer Institut, CH-5232 Villigen PSI, Switzerland}

\author{R.~Khasanov}
\affiliation{Laboratory for Muon Spin Spectroscopy, Paul Scherrer Institut, CH-5232 Villigen PSI, Switzerland}

\author{H.~Keller}
\affiliation{Physik-Institut der Universit\"{a}t Z\"{u}rich, Winterthurerstrasse 190, CH-8057 Z\"{u}rich, Switzerland}

\begin{abstract}
An extended investigation of the electronic phase diagram of FeSe$_{1-x}$ up to pressures of $p\simeq2.4$\,GPa by means of ac and dc magnetization, zero field muon spin rotation (ZF $\mu$SR), and neutron diffraction is presented. ZF $\mu$SR indicates that at pressures $p\geq0.8$\,GPa static magnetic order occurs in FeSe$_{1-x}$ and occupies the full sample volume for $p\gtrsim 1.2$\,GPa. ac magnetization measurements reveal that the superconducting volume fraction stays close to $100$\% up to the highest pressure investigated. In addition, above $p\geq1.2$\,GPa both the superconducting transition temperature $T_{\rm c}$ and the magnetic ordering temperature $T_{\rm N}$ increase simultaneously, and both superconductivity and magnetism are stabilized with increasing pressure. Calculations indicate only one possible muon stopping site in FeSe$_{1-x}$, located on the line connecting the Se atoms along the $c$-direction. Different magnetic structures are proposed and checked by combining the muon stopping calculations with a symmetry analysis, leading to a similar structure as in the LaFeAsO family of Fe-based superconductors. Furthermore, it is shown that the magnetic moment is pressure dependent and with a rather small value of $\mu\approx 0.2\,\mu_B$ at $p\simeq2.4$\,GPa. 
\end{abstract}


\maketitle

\section{Introduction}
Shortly after the discovery of superconductivity in the Fe-based compound LaFeAsO$_{1-x}$F$_y$ in 2008 by Kamihara \textit{et al.},\cite{Kamihara_JACS_08} Hsu \textit{et al.}\cite{Hsu_PNAS_08} observed superconductivity in the basic binary compound FeSe$_{1-x}$. This simple system shares the superconducting layers consisting of a Fe square planar sheet tetrahedrally coordinated by As/P or Se/Te atoms as a common feature with all of the Fe-based superconductors. Most of the known Fe-based superconductors are made up of a stack of the electronically active layers, separated by layers that act as a charge reservoir to dope the Fe-As/Se layers. FeSe$_{1-x}$ is an exception to that rule because it consists of a stack of superconducting layers only. In this binary system the superconducting transition temperature is $T_{\rm c}\simeq8$\,K. Thus, it could be argued that this is more a conventional superconductor.\cite{Hsu_PNAS_08} Shortly after, the electronic and magnetic phase diagram under pressure was studied.\cite{Medvedev_Nat_09,Margadonna_PRB_09} It was found that the transition temperature exhibits one of the largest pressure effects on $T_{\rm c}$ known. It reaches values of $T_{\rm c}\approx 37$\,K at $p\approx9$\,GPa, demonstrating that FeSe$_{1-x}$ in fact is a high temperature superconductor. 
Furthermore, it was found that tetragonal FeSe$_{1-x}$ undergoes a structural phase transition starting at $p\sim9$\,GPa from a tetragonal to a hexagonal, non-superconducting and more densely packed phase. With increasing pressure the volume fraction of the tetragonal phase as well as $T_{\rm c}$ decrease until at high pressures ($p\geq20$\,GPa) only the non-superconducting hexagonal phase is present.\cite{Margadonna_PRB_09} 
Early muon spin rotation ($\mu$SR) experiments on FeSe$_{1-x}$ revealed that the system is non-magnetic at ambient pressure down to $T=0.02$\,K.\cite{Khasanov_PRB_08} The investigation of the pressure dependence also did not show magnetic order in the beginning up to the highest pressures, just before the structural phase transition occurs.\cite{Medvedev_Nat_09} This is in striking contrast to the other Fe-based superconductors that usually exhibit static magnetic order in the parent compound. This is unexpected, since the FeSe$_{1-x}$ layers are isoelectric to those of the parent compounds of other Fe-based superconductors.\cite{Buchner_NAT_09} Shortly after, however, NMR studies showed a wipeout of the signal that revealed an incipient magnetic phase transition under pressure.\cite{Imai_PRL_09} This possibly may be interpreted as static magnetic order with a broad field distribution or as slow spin fluctuations, since no magnetic order was observed by $\mu$SR at ambient pressure. It seems that both the magnetic and the superconducting states stabilize with increasing pressure. 
In fact, static magnetic ordering was observed above $p\sim0.8$\,GPa by means of $\mu$SR.\cite{Bendele_PRL_10} The experiments revealed that as soon as magnetic ordering occurs, the magnetic and the superconducting states seem to compete with each other. This is because the incommensurate magnetic order gets suppressed when superconductivity sets in and, in addition, $T_{\rm c}$ decreases in the pressure region $0.8\leq p\leq1.2$\,GPa. Above $p\simeq1.2$\,GPa both ground states apparently coexist on an atomic length scale. Both the magnetic ordering temperature $T_{\rm N}$ and $T_{\rm c}$ increase simultaneously with increasing pressure, and the magnetic order becomes commensurate.\cite{Bendele_PRL_10}

In this paper an extended study of the electronic and magnetic properties of FeSe$_{1-x}$ under pressure investigated by means of ac susceptibility and $\mu$SR is presented. In addition, magnetic structures of FeSe$_{1-x}$ under pressure are proposed and checked by neutron diffraction measurements. The magnetic moment in the ordered state is estimated for different pressures. Furthermore, the discrepancy between M\"ossbauer\cite{Medvedev_Nat_09} and $\mu$SR results\cite{Bendele_PRL_10} is discussed under the aspect that the samples used in each study were prepared by slightly different methods.\cite{Pomjakushina_09_PRB,McQueen_09_PRB} 

\section{samples}

\begin{figure}[t!]
\centering
\vspace{-0cm}
\includegraphics[width=1\linewidth]{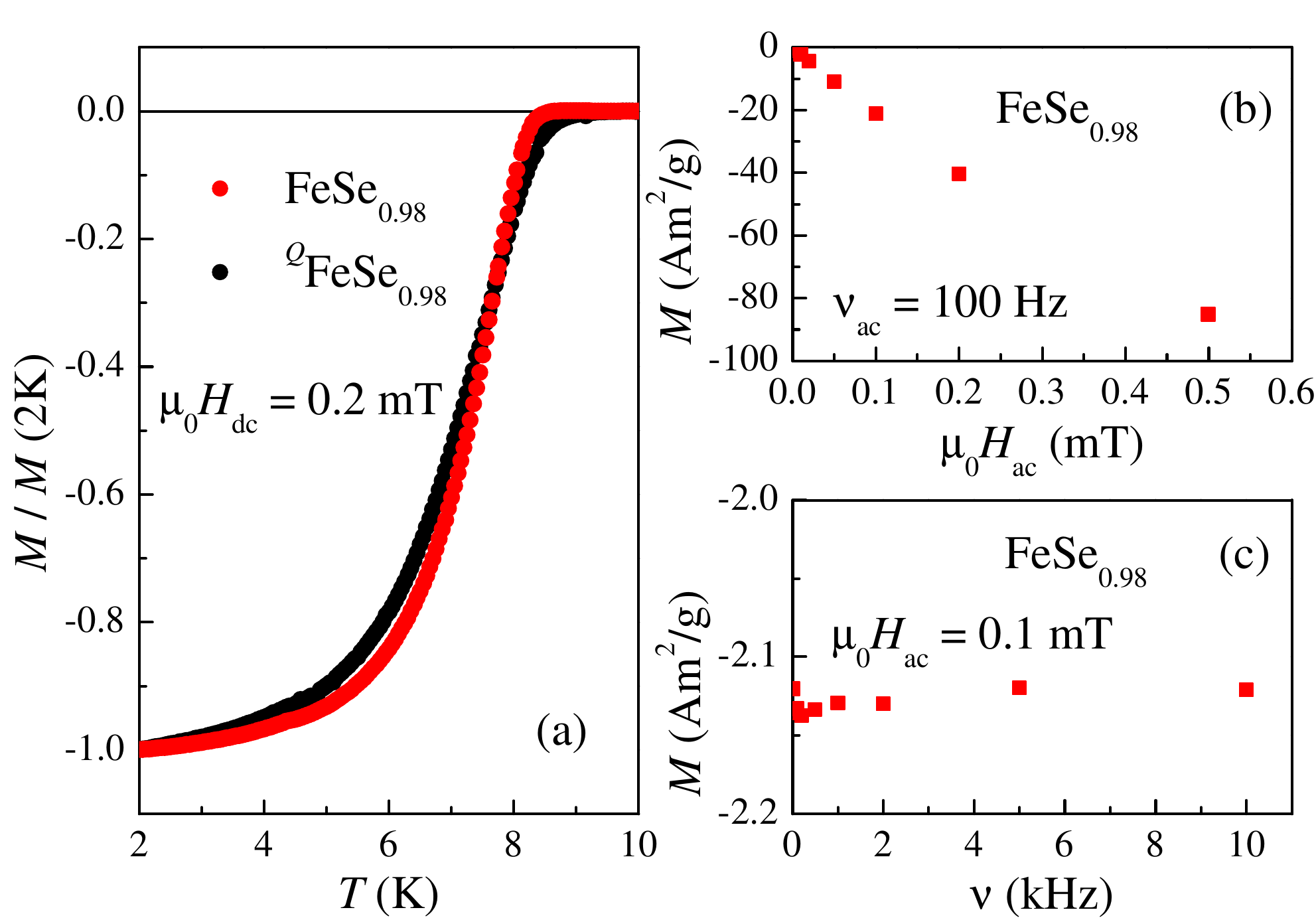}
\caption{(color online) (a) Temperature dependence of the normalized magnetization $M/M(2\text{\,K})$ for FeSe$_{0.98}$ and $^Q$FeSe$_{0.98}$ (see text for details on sample notation) at $p=0$\,GPa. The transition temperature $T_{\rm c}$ of both samples, obtained by the intersection of straight lines fit to the data above and below the transition is $\simeq8$\,K. The shapes of the magnetization curves for the two samples are very similar. (b) Dependence of the ac magnetization $M_{\rm ac}$ on the ac field amplitude $\mu_0H_{\rm ac}$ at a fixed frequency $\nu_{\rm ac}=100$\,Hz (c) and the ac frequency $\nu_{\rm ac}$ at a fixed ac field amplitude $\mu_0 H_{\rm ac} = 0.1 $\,mT. }
\label{fig_magn1}
\end{figure}

The FeSe$_{1-x}$ samples were prepared following the procedures described in Refs.~\onlinecite{Pomjakushina_09_PRB} and \onlinecite{McQueen_09_PRB}. In both methods the samples are placed in sealed silica tubes and are prepared in two steps. In the first step Pomjakushina \textit{et al.}\cite{Pomjakushina_09_PRB} used selenium and iron powders as starting materials and synthesized FeSe$_{1-x}$ in a solid state reaction at temperatures ranging from $400-700^\circ$C. After powderizing the samples in He-atmosphere, they were reannealed at $700\,^\circ$C, then the temperature was stabilized at $420\,^\circ$C, and finally they were cooled slowly to room temperature. McQueen \textit{et al.},\cite{McQueen_09_PRB} on the other hand, used shots of selenium iron pieces. They were molten at $1075\,^\circ$C, powderized and annealed again at $T\sim400\,^\circ$C. However, the main difference of the two procedures is that the samples prepared by the method of McQueen \textit{et al.}\cite{McQueen_09_PRB} are quenched from $\sim400\,^\circ$C to $\sim -15\,^\circ$C, whereas the samples prepared after Pomjakushina \textit{et al.}\cite{Pomjakushina_09_PRB} are cooled slowly from $\sim 400\,^\circ$C to room temperature. Here, the specimens are denoted as FeSe$_{0.98}$ for the slowly cooled ones, and $^Q$FeSe$_{0.98}$ for the quenched ones. All samples were found to be phase pure with a superconducting transition temperature of $T_{\rm c}\simeq 8$\,K. In fact, the transitions to the superconducting state is for both preparation procedures very similar (see Fig.~\ref{fig_magn1}a).\cite{Ichsanow_2010}

\section{superconducting properties}

The superconducting properties of FeSe$_{1-x}$ were studied by means of ac and dc magnetization measurements (Fig.~\ref{fig_magn1}). The zero field cooled dc measurements, preformed in a commercial \textit{Quantum Design MPMS} SQUID 7\,T magnetomenter in $\mu_0H=0.2$\,mT, revealed $T_{\rm c}\simeq8$\,K for both FeSe$_{0.98}$ and $^Q$FeSe$_{0.98}$. The ac magnetization measurements under pressure were performed in a home made ac susceptometer in piston-cylinder pressure cells, especially designed for $\mu$SR experiments. The ac amplitude was $\mu_0H_{\rm ac}\approx 0.1$\,mT and the frequency was $\nu_{\rm ac}=94$\,Hz. As a pressure transmitting medium 7373 Daphne oil was used. The pressure applied was measured in situ by monitoring the the shift of $T_{\rm c}$ of Pb or/and In. To ensure that the position of the sample in the cell is the same for all pressures investigated the pick up and excitation coils were directly wound on the pressure cell.
Additional ac magnetization measurements were performed to check whether the ac signal under pressure was entirely determined by the bulk Meissner response of each grain. Thus, other effects like e.g. weak links between the individual grains or surface superconductivity can be excluded. This was done on a commercial \textit{Quantum Design PPMS} in various fields ($0\geq\mu_0H_{\rm AC}\geq0.5$\,mT) and frequencies ($0\geq\nu\geq599$\,Hz). 
As shown in Fig.~\ref{fig_magn1}b and c the experiments reveal that the ac magnetization scales linearly with the field and is independent of frequency as expected for a superconductor in the Meissner state. 

\begin{figure}[t!]
\centering
\vspace{-0cm}
\includegraphics[width=1\linewidth]{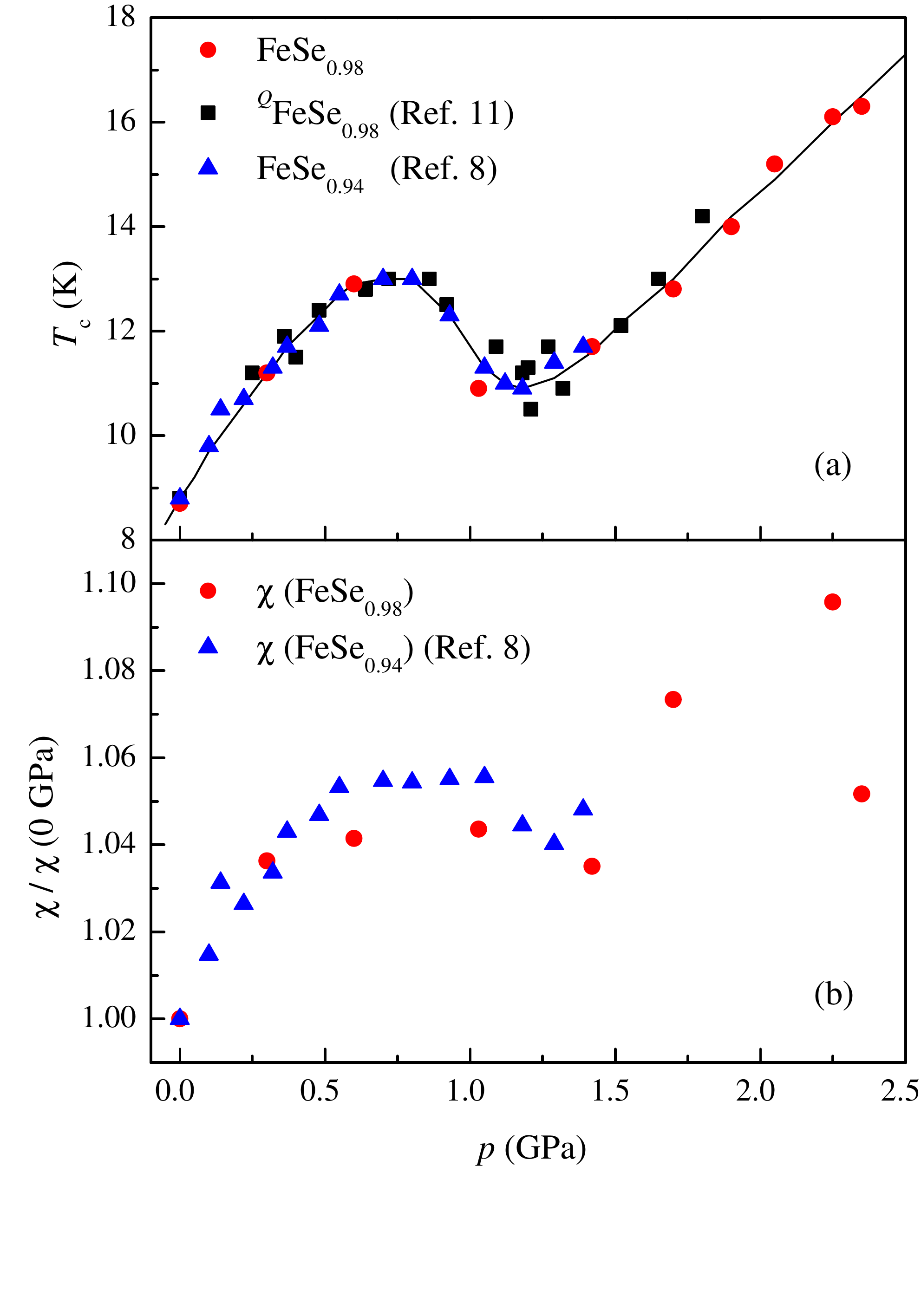}\vspace{-1cm}
\caption{(color online) (a) Dependence of the superconducting transition temperature $T_{\rm c}$ on pressure $p$ of FeSe$_{1-x}$. The line is a guide to the eyes. (b) Pressure $p$ dependence of the ac susceptibility $\chi$ normalized to the ambient pressure value $\chi(0\,{\rm GPa)}$ at $T=6$\,K, indicating bulk superconductivity for all pressures investigated. See text for details on sample notation.}
\label{fig_magn2}
\end{figure}

The superconducting transition temperature of FeSe$_{1-x}$ (FeSe$_{0.98}$ and $^Q$FeSe$_{0.98}$) is $T_{\rm c}\simeq 8$\,K at ambient pressure (see Fig.~\ref{fig_magn1}). Upon applying hydrostatic pressure FeSe$_{1-x}$ exhibits one of the highest pressure effects known on $T_{\rm c}$. The overall increase of $T_{\rm c}$ is non monotonic and shows a local maximum at $p\simeq0.8$\,GPa, followed by a local minimum at $p\simeq1.2$\,GPa (Fig.~\ref{fig_magn2}a). This behavior is similar to that already observed earlier both by dc and ac magnetization.\cite{Bendele_PRL_10,Miyoshi_JPSJ_09,Masaki_JPSJ_09} In the region where $T_{\rm c}$ decreases static magnetism develops in the sample and competes with superconductivity (see below and Ref.~\onlinecite{Bendele_PRL_10}). Upon increasing the pressure above $p=1.2$\,GPa the superconducting transition temperature increases again and reaches values of $\sim16$\,K at the highest pressure investigated in this study ($2.4$\,GPa). 

In Fig.~\ref{fig_magn2}b the diamagnetic response at $T=6$\,K normalized to the value at ambient pressure is shown as a function of pressure. At ambient pressure the value of the measured ac voltage of the samples in the pressure cell is equal to the magnetization of the sample measured in the PPMS magnetomenter without a pressure cell. Thus, the ac voltage in the pressure cell is representing the superconducting response of FeSe$_{1-x}$. Calculating the susceptibility from the magnetization measurements allows to estimate the superconducting volume fraction. The susceptibility was determined to $\chi_{\rm ac}\simeq 1.3$ (Fig.~\ref{fig_magn1}a). By assuming the samples consist of individual sphere like shaped grains with a demagnetization factor of $n\simeq1/3$ leads to an ideal diamagnetic response of $\chi=-1$. This indicates that FeSe$_{1-x}$ is a bulk superconductor with a superconducting volume fraction close to 100\%. Since the absolute value of the ac response measured at 6\,K for each individual pressure is similar, it is concluded that the sample is a bulk superconductor up to the highest pressure investigated.

\section{Magnetic properties}
The magnetic response of FeSe$_{1-x}$ for various pressures was studied by means of zero-field muon spin rotation experiments (ZF $\mu$SR). The experiments were carried out using the $\mu$E1 beam line at the GPD instrument at the Paul Scherrer Institute (PSI, Switzerland) at temperatures ranging from $0.25$ to $80$\,K. The $\mu$SR time spectra were analyzed using the free software package MUSRFIT.\cite{MUSRFIT} ZF $\mu$SR is a well known technique to study magnetically ordered phases where the muon acts as a local magnetic microprobe. Positively charged muons are implanted into the sample where they thermalize after a short time ($<10^{-13}$\,s). Once stopped at an interstitial site the muon interacts with its local environment and decays after its lifetime of $\tau_\mu=2.197\,\mu$s into a positron and two neutrinos. The positron is emitted predominantly along the muon spin direction at the time of decay. Thus, by monitoring the time evolution of the muon spin polarization, information on the local magnetic field at the muon stopping site $B_{\rm int}$ and the magnetic volume fraction are obtained. 

The $\mu$SR signal in a pressure cell consists of a superposition of two components, one arising from the sample ($\mathcal A^{\rm S}$) and one from the pressure cell ($\mathcal A^{\rm PC}$): 
\begin{equation}
 \mathcal A(t)=\mathcal A^{\rm PC}(t)+\mathcal A^{\rm S}(t)
\label{eq_1}
\end{equation}
In the data analysis the ratio of the component of the pressure cell and the component of the sample $\mathcal A^{\rm PC}/\mathcal A^{\rm S}$ was kept constant for each individual pressure and was always $\approx 50$\,\%. For the present study two different pressure cells consisting of MP35N and CuBe were used. The ZF response of the empty cells is described elsewhere.\cite{Andreica}

\begin{figure}[t!]
\centering
\vspace{-0cm}
\includegraphics[width=1\linewidth]{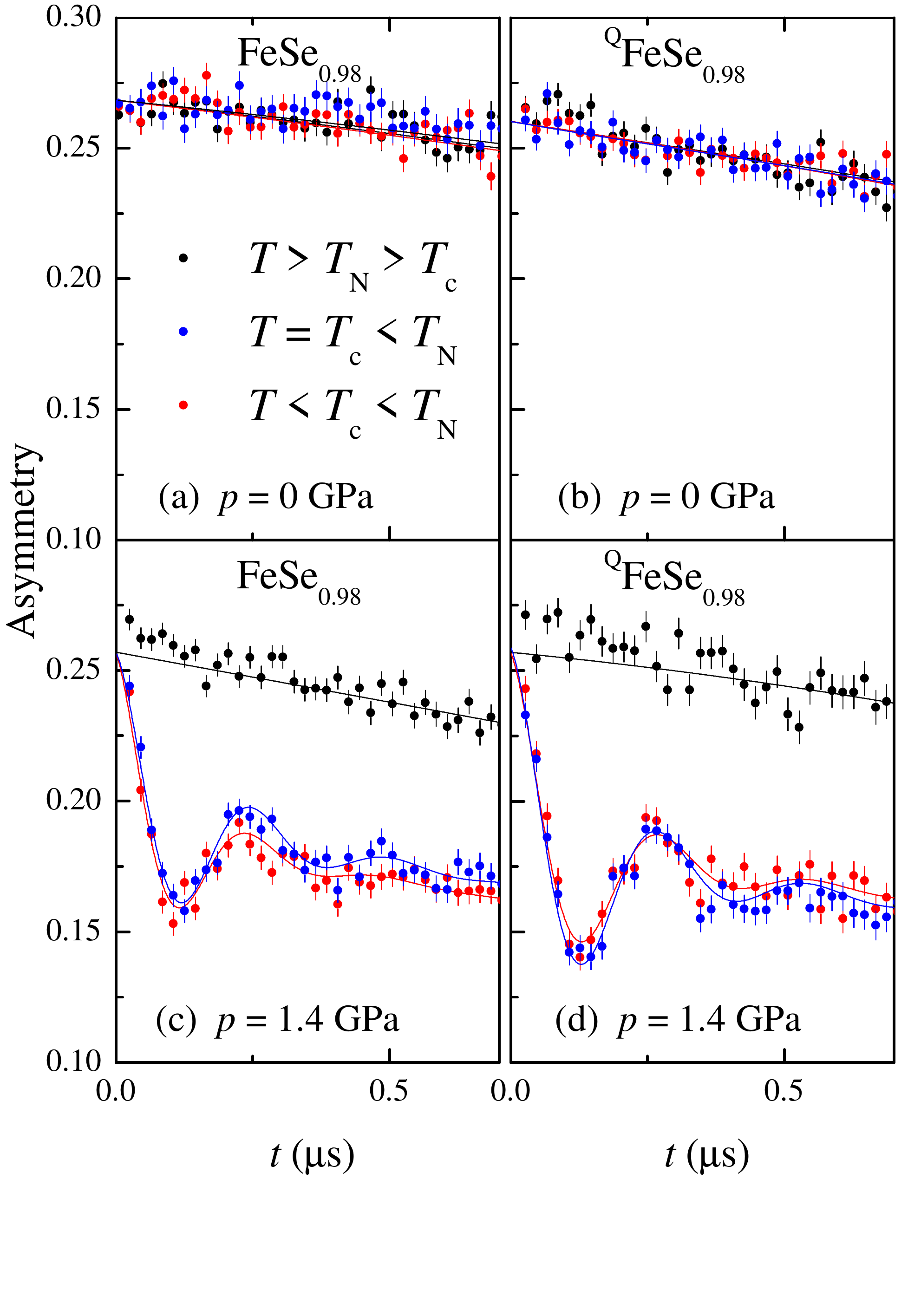}
\vspace{-0.5cm}
\caption{(color online). Zero-field $\mu$SR time spectra of FeSe$_{0.98}$ for (a) $p=0$ and (c) $p=1.4$\,GPa, and $^Q$FeSe$_{1-x}$ for (b) $p=0$ and (d) $p=1.4$\,GPa for different temperatures. The lines are fits of Eq.~(\ref{eq_1}) to the data. }
\label{fig_musr_raw}
\end{figure}

As we reported earlier,\cite{Bendele_PRL_10} in the low pressure region, where $T_{\rm c}$ increases linearly with $p$, no magnetic order is observed in all of the samples. The $\mu$SR time spectra are overlapping for all temperatures, indicating the same magnetic state for all temperatures measured (Fig.~\ref{fig_musr_raw}a and b). The $\mu$SR time spectra were analyzed using a single exponential decay function:
\begin{equation}
 \mathcal A^{\rm S}(t)=A^{\rm S}_0\exp[-\Lambda_0t]
\end{equation}
Here $\Lambda_0$ is the Lorentzian depolarization rate. The exponential behavior at low pressures indicates the presence of diluted and randomly distributed and oriented magnetic moments in the sample volume which can be attributed to traces of Fe impurities.\cite{Khasanov_PRB_08}

As shown in Fig.~\ref{fig_musr_raw}c and d for $p\gtrsim0.8$\,GPa spontaneous muon-spin precession is observed, reflecting the appearance of static magnetic order below the N\'eel temperature $T_{\rm N}$. The analysis was made by taking into account that the magnetic order appears gradually: one part of the muons experiences a static local field and the other part stops in non-magnetic regions:
\begin{align}
\nonumber \mathcal A^{\rm S}(t)=&A^{\rm S}_0\Bigg{(}m\left(\frac{2}{3}f_{\rm osc}\exp[-\Lambda_{\rm t}t]+ \frac{1}{3}\exp[-\Lambda_{\rm l}t]\right)\\&+\left(1-m\right)\exp[-\Lambda_0t]\Bigg)
\end{align}
Here $m$ is the magnetic volume fraction of the sample, $f_{\rm osc}$ represents the magnetic signal of the sample and has, depending on pressure, the form $f_{\rm osc}=\cos(\omega_0t+\phi_0)$ or $f_{\rm osc}=j_0(\omega_0t+\phi_0)$, whereas $\omega_0$ is the precession frequency, $j_0$ is a zeroth-order spherical Bessel function, and $\phi_0$ the initial phase of the muon ensemble. The parameters $\Lambda_{\rm t}$ and $\Lambda_{\rm l}$ describe the relaxation transverse and longitudinal to the muon spin of the magnetic signal, respectively.

\begin{figure}[t!]
\centering
\vspace{-0cm}
\includegraphics[width=1\linewidth]{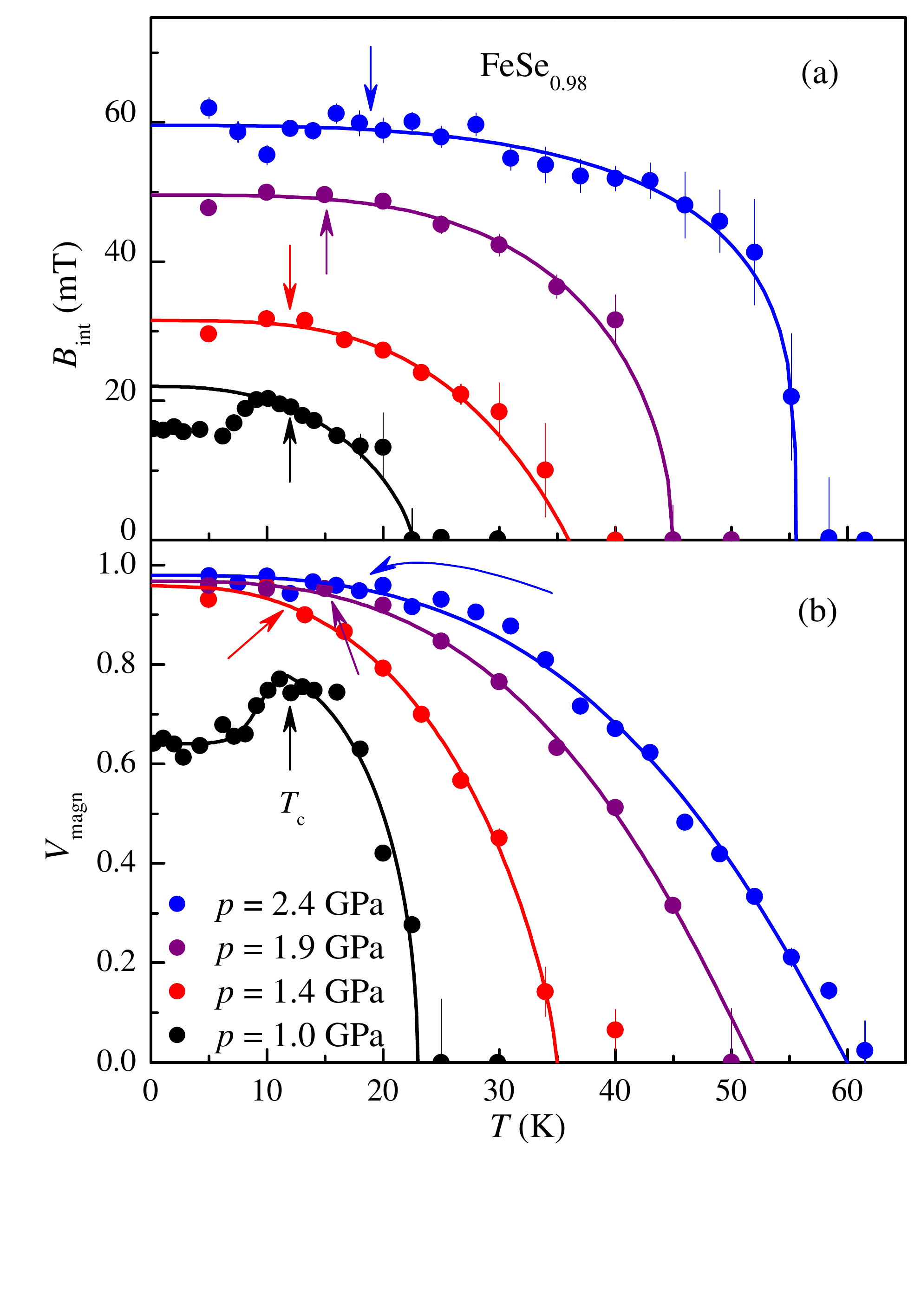}\vspace{-1cm}
\caption{(color online) Temperature dependence of (a) the internal magnetic field at the muon stopping site $B_{\rm int}$ and of (b) the magnetic volume fraction for FeSe$_{1-x}$ for various pressures. Both parameters are obtained directly from the fit of Eq.~(\ref{eq_1}) to the data. The solid lines in (a) correspond to fits of $B_{\rm int}(T)$ in the region $T_{\rm c}\geq T\geq T_{\rm N}$ to Eq.~(\ref{eq_BT}). For details see text. The solid lines in (b) are a guide to the eyes. The arrows indicate the superconducting transition temperature $T_{\rm c}$. }
\label{fig_musr_data}
\end{figure}

In the pressure region where $T_{\rm c}$ decreases, both the magnetic and superconducting ground state are competing. This is seen first by the decrease of $T_{\rm c}$ (Fig.~\ref{fig_magn2}a) and second by a decrease of the frequency and the magnetic volume fraction $m$ below $T_{\rm c}$ (see Fig.~\ref{fig_musr_data}a and b, and Ref.~\onlinecite{Bendele_PRL_10}). In this region the magnetic signal is described best by a Bessel function which indicates the presence of incommensurate magnetic order in the samples.\cite{Savici_PRB_02} 

As shown in Fig.~\ref{fig_musr_data} for $p\gtrsim1.2$\,GPa (where the local minimum of $T_{\rm c}$ is reached) superconductivity and magnetic order coexist in the full sample volume. Here, the magnetic volume fraction reaches $100$\% and stays constant in the superconducting state down to the lowest temperature where also the superconducting volume fraction remains constant at $\simeq100$\% (see Fig.~\ref{fig_magn2}). Moreover, $B_{\rm int}$ is not significantly changing (decreasing) below $T_{\rm c}$, and the magnetic order changes from an incommensurate to a commensurate as reflected in the $\mu$SR line shape which is described better by a damped cosine function with zero initial phase than by a Bessel function. This indicates coexistance of superconductivity and magnetism in the full sample voulume. 

\begin{figure}[t!]
\centering
\vspace{-0cm}
\includegraphics[width=1\linewidth]{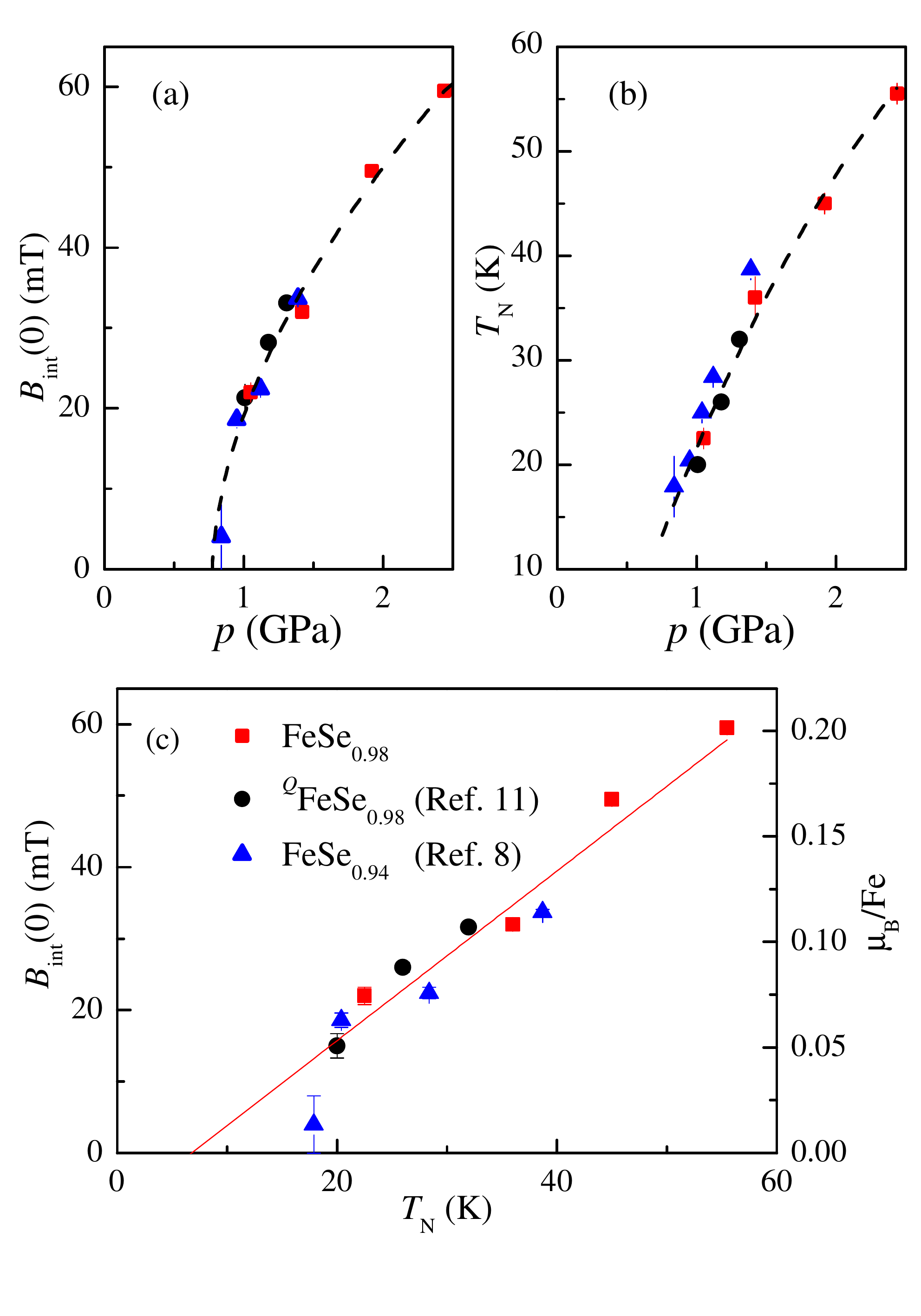}\vspace{-0.5cm}
\caption{(color online) (a) Pressure dependence of the internal magnetic field at the muon stopping site $B_{\rm int}(0)$. (b) Pressure dependence of the magnetic ordering temperature $T_{\rm N}$. The dotted lines in (a) and (b) are guides to the eye. The dependence of $B_{\rm int}$ on $T_{\rm N}$ is shown (c). See text for details on sample notation.}
\label{fig_musr_complete}
\end{figure}

To determine the zero-temperature value of $B_{\rm int}(0)$ and $T_{\rm N}$ the temperature dependence of $B_{\rm int}(T)$ was fitted to the power law expression:
\begin{equation}
 B_{\rm int}(T)=B_{\rm int}(0)\left(1-\left(\frac{T}{T_{\rm N}}\right)^\alpha\right)^\beta.
\label{eq_BT}
\end{equation}
Here $\alpha$ and $\beta$ are the power exponents. For the pressure region in which $B_{\rm int}$ decreases in the superconducting state, only the data above $T_{\rm c}$ were used to analyze the data with Eq.~(\ref{eq_BT}). The obtained values of $B_{\rm int}(0)$ and $T_{\rm N}$ are plotted in  Fig.~\ref{fig_musr_complete}a and b together with the results from earlier studies of FeSe$_{0.94}$ and $^Q$FeSe$_{0.98}$.\cite{Bendele_PRL_10,Ichsanow_2010} For all samples $B_{\rm int}(0)$ increases with increasing pressure (see Fig.~\ref{fig_musr_complete}a). As shown in Fig.~\ref{fig_musr_complete}b the N\'eel temperature increases in parallel from $T_{\rm N}=17$\,K at $p=0.8$\,GPa where magnetism appears in FeSe$_{1-x}$ with increasing pressure to $T_{\rm N}=55$\,K at the maximum pressure $p\simeq2.4$\,GPa investigated here. No tendency for a saturation at high pressures of both $B_{\rm int}(0)$ and $T_{\rm N}$ is observed. 

Unlike the $\mu$SR experiments presented here, an earlier M\"ossbauer study did not reveal magnetic order under pressure in FeSe$_{1-x}$.\cite{Medvedev_Nat_09} However, the samples used in this study were prepared after the method proposed by McQueen \textit{et al.}\cite{McQueen_09_PRB} As mentioned already above, samples denoted as $^Q$FeSe$_{1-x}$ were prepared following exactly the recipe of McQueen \textit{et al.}\cite{McQueen_09_PRB}~and were investigated by means of $\mu$SR.\cite{Ichsanow_2010} 
In contrast to the earlier study of Ref.~\onlinecite{McQueen_09_PRB} they also show a similar magnetic behavior as the samples prepared by our method (see Fig.~\ref{fig_musr_raw}b and d). In particular they also show magnetic order upon applying pressure. A simple explanation of this discrepancy could be that magnetism was overseen. Low temperature ($T=4.2$\,K) M\"ossbauer spectra were taken only at few pressures: At ambient pressure no magnetic order in agreement with the $\mu$SR experiments was seen, and at $p=14.4$\,GPa and 19.7\,GPa no magnetic hyperfine splitting in the M\"ossbauer spectra was observed.

In Fig.~\ref{fig_musr_complete}c $B_{\rm int}(0)$ vs.~$T_{\rm N}$ is plotted, indicating that the magnetic moment is increasing with increasing $T_{\rm N}$. This points to a more robust magnetic order with increasing pressure.
When the magnetic order is fully established (above $p=1.2$\,GPa; the magnetic volume fraction reaches $100$\%) $T_{\rm c}$ starts to increase again (Fig.~\ref{fig_magn2}a) simultaneously with $T_{\rm N}$ up to $T_{\rm N}\approx 60$\,K and $T_{\rm c}\approx 16$\,K at the highest investigated pressure in this study. 
It seems that both order parameters are stabilized at high pressures: (i) Both $T_{\rm c}$ and $T_{\rm N}$ increase with increasing pressure, (ii) the magnetic and superconducting volume fractions stay $100$\,\% even below $T_{\rm c}$ to the highest investigated pressure, and (iii) the internal magnetic field $B_{\rm int}(0)$ increases with increasing pressure for all samples studied. 

\section{Muon Stopping Site and magnetic moment}

Up to now it is not clear what kind of magnetic structure develops in FeSe$_{1-x}$ under pressure. Calculations of the muon stopping sites at different pressures were performed and combined with a symmetry analysis to check for possible different magnetic structures.

The space group symmetry of FeSe$_{1-x}$ at low temperatures is Cmma with Fe in the $4a$-position $(1/4,0,0)$ and Se in the $4g$-position $(0,1/4,z)$ (see for instance Ref.~\onlinecite{Louca_PRB_10}). Here the symmetry of the FeSe$_{1-x}$ layers exactly resembles the symmetry of the FeAs-layers in the LaFeAsO compound with the same Cmma space group which remains unchanged in FeSe$_{1-x}$ up to a pressure $p\approx9$\,GPa.\cite{Margadonna_PRB_09}

In order to evaluate possible muon sites the modified Thomas Fermi approach\cite{Reznik_PRB_95} and available structural data were used.\cite{Margadonna_PRB_09} This method allows to determine directly the self consistent distribution of the valent electron density from which the electrostatic potential is obtained. Local interstitial minima of this potential serve as stopping sites for muons. The applicability of this approach was verified by comparing the numerical results with the experimentally determined muon sites in $R$FeO$_3$\cite{Holzschuh_PRB_83} ($R=$ rare earth) and by a successful interpretation of $\mu$SR spectra of the complex magnetic structures in layered cobaltites $R$BaCo$_2$O$_{5.5}$\cite{Luetkens_PRL_08} and Fe-pnictides $R$FeAsO.\cite{Maeter_PRB_09}

\begin{figure}[t!]
\centering
\vspace{-0cm}
\includegraphics[width=\linewidth]{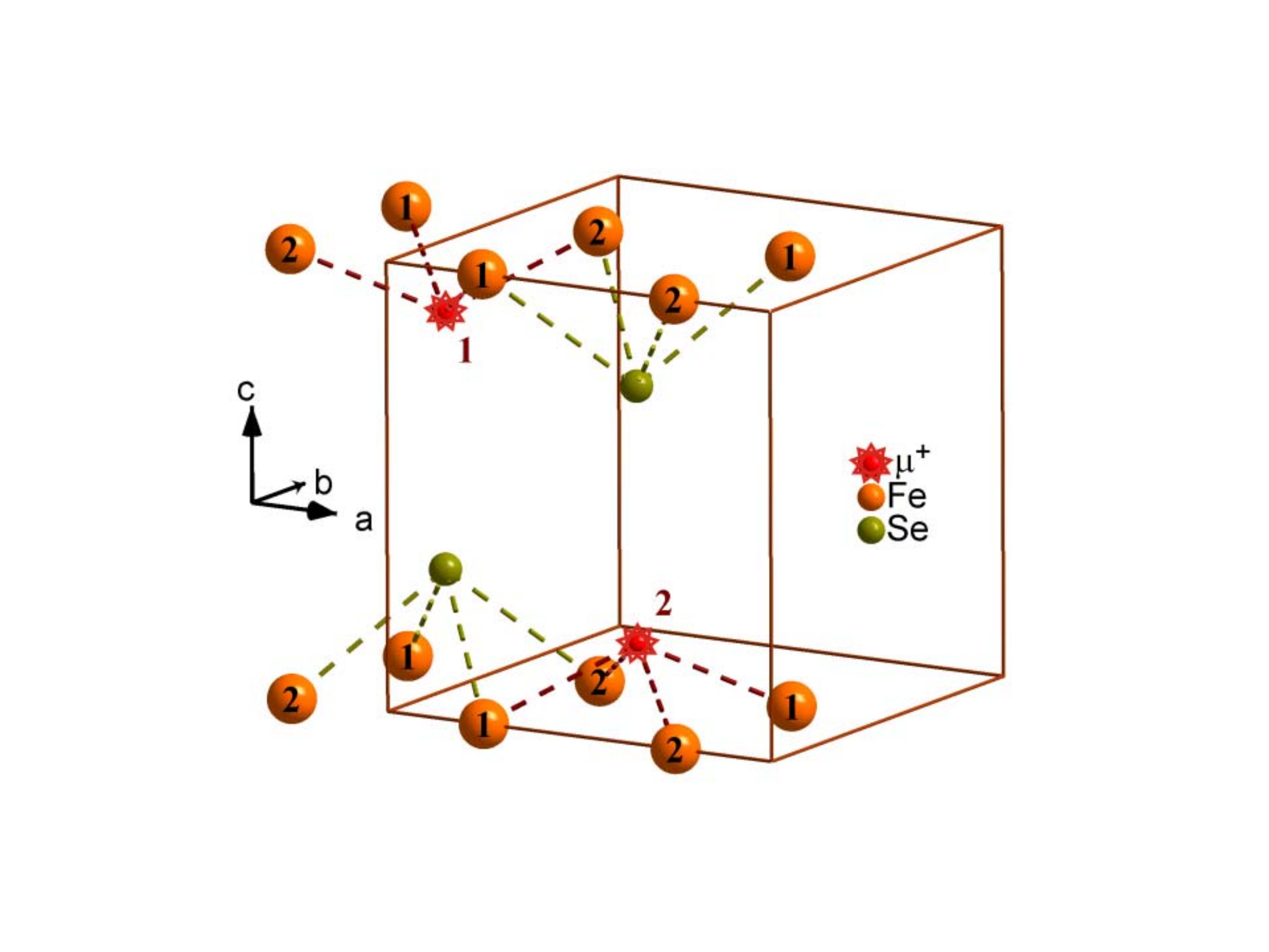}
\caption{(color online) 
The crystallographic unit cell of FeSe$_{1-x}$ in the Cmma setting. The enumeration of the Fe atoms and the muon positions is shown.  }
\label{fig_muon_crystal}
\end{figure}

Only one possible muon stopping site is observed. It is located on the line connecting the Se - Se ions along the $c$-direction with the coordinates $(0, 1/4, z)$ and has the $4g$ local point symmetry (mm2) \textit{i.e.}~the same as the Se ions. The position of the muon sites in the crystallographic cell is shown in Fig.~\ref{fig_muon_crystal}. Note that the crystallographic unit cell differs from the primitive cell which is built by primitive translations $a_1=(a/2,b/2,0)=(\tau_x,\tau_y,0)$, $a_2=(-a/2,b/2,0)=(-\tau_x,\tau_y,0)$, and $a_3=(0,0,c)=(0,0,2\tau_z)$.

As seen in Table~\ref{tab_crystal}, application of pressure leads to a general increase of the distance of the calculated muon stopping sites to the iron $ab$-plane, whereas the angles of the Fe-Se-Fe bonds $\alpha_a$ (along the $a$-direction) and $\alpha_b$ (along the $b$-direction) are almost identical at ambient pressure. However, at higher pressures they tend to differ.

\begin{table}[t]\centering\caption{The pressure dependence of the calculated muon position and Fe-Se-Fe bond angles $\alpha_a$ (along the $a$-direction) and $\alpha_b$ (along the $b$-direction). The crystallographic data are from Refs.~\onlinecite{Margadonna_PRB_09} and \onlinecite{Louca_PRB_10}. }\hspace{20mm}
 \begin{tabular}{ c  c  m{18mm}   m{18mm}   c } \toprule
  $p$~(GPa) & $T$~(K) & \multicolumn{2}{c }{Fe-Se-Fe bond angle} & \parbox{2cm}{$z$-coordinate\\ of 4g muon site}\\ \hline
 & & \centering$\alpha_a$ &\centering $\alpha_b$ &\\ \hline
 $0$\footnotemark[1] & 7 & \centering$  67.781$&\centering$67.551$&0.84\\ 
 $0.25$\footnotemark[2] & 16 & \centering$67.920$&\centering$68.086$&0.84\\
 $4.0$\footnotemark[2] & 16 & \centering$67.688$&\centering$68.216$&0.83\\ 
 $9.0$\footnotemark[2] & 16 & \centering$67.097$&\centering$67.532$&0.81\\ \toprule
 \end{tabular}
 \footnotetext[1]{Louca \textit{et al.}\cite{Louca_PRB_10}}\footnotetext[2]{Margadonna \textit{et al.}\cite{Margadonna_PRB_09}}
\label{tab_crystal}
\end{table}

The stronger reduction of the $c$-axis compared to the $a$- and $b$-axis leads to an increase of the Fe-Se-Fe bond angle that can be interpreted as a tendency to antiferromagnetic exchange in accordance with the semi empirical Goodenough Kanamori rules.\cite{ Anderson_PR_1950,Goodenough_PR_1955,Kanamori_JPCS_1959} Note that already small variations of the Fe-As-Fe bond angles along $a$- and $b$-axes in the $R$FeAsO compounds lead to a drastic change of the magnetic exchange sign from anti-ferromagnetic (positive) along $a$-axis to ferromagnetic (negative).\cite{Han_PRL_2009} 
However, opposite to the $R$FeAsO the $b$-axis remains in FeSe$_{1-x}$ larger than the $a$-axis for all pressures. Due to this similarity one can suppose the occurrence of a ferromagnetic type of order along the $a$-axis and an antiferromagnetic one along the $b$-axis in FeSe$_{1-x}$ under pressure. The minimal model which could account for this feature should include a doubling of the primitive cell along the $b$-axis with magnetic propagation vectors either $K_{I}=(0,\pi/\tau_y,\pi/2\tau_z)$ or $K_{II}=(0,\pi/\tau_y,0)$. Additionally, more simple possible magnetic vectors such as $K_0=(0,0,0)$ and $K_{III}=(0,0,\pi/2\tau_z)$ are considered. 

\begin{figure}[t!]
\vspace{4cm}
\hspace{-0cm}\includegraphics[width=1.3\linewidth]{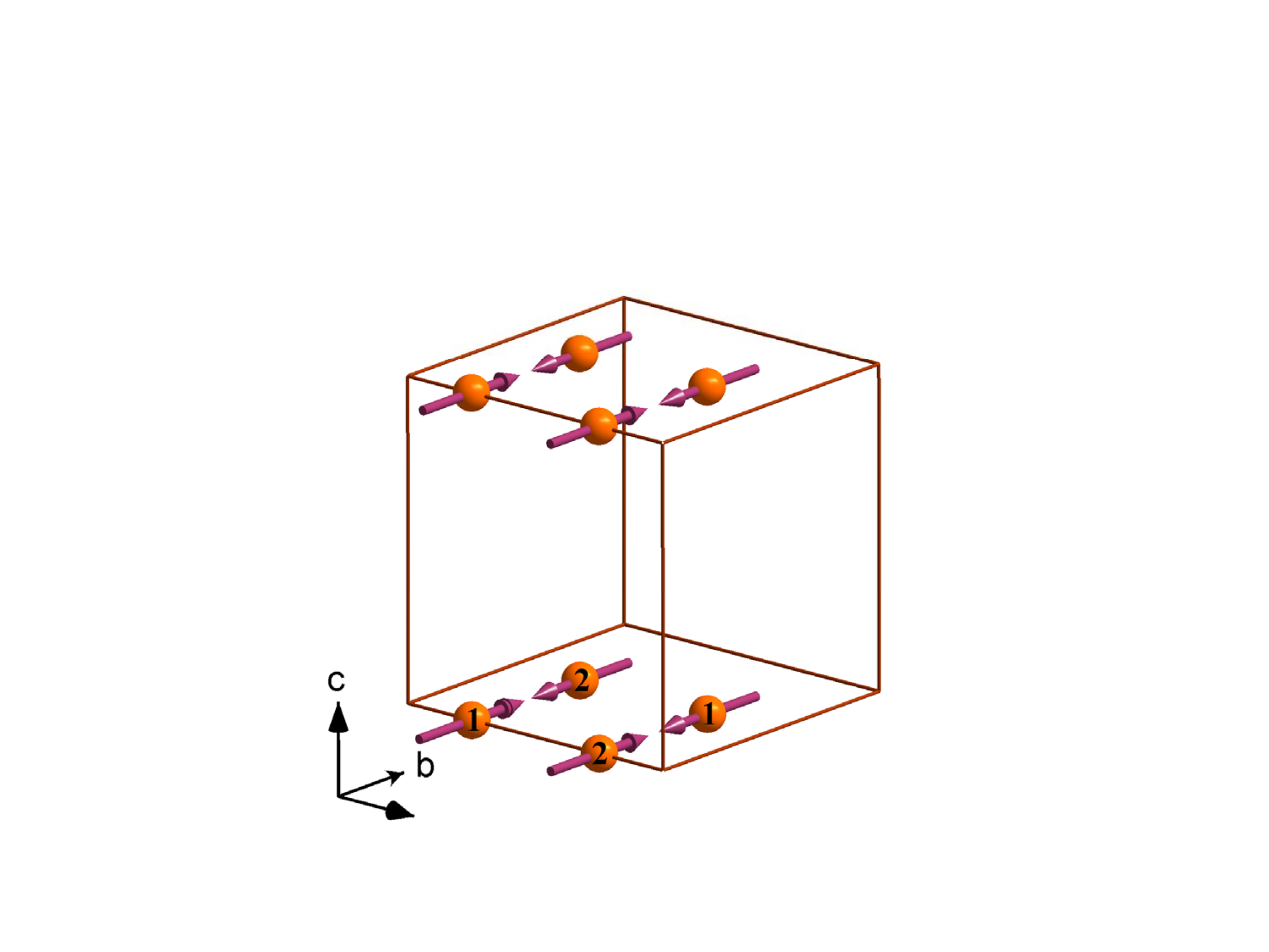}\\
\vspace{-6cm}
\flushleft (b)\\ 
\vspace{-9cm}
\includegraphics[width=1.3\linewidth]{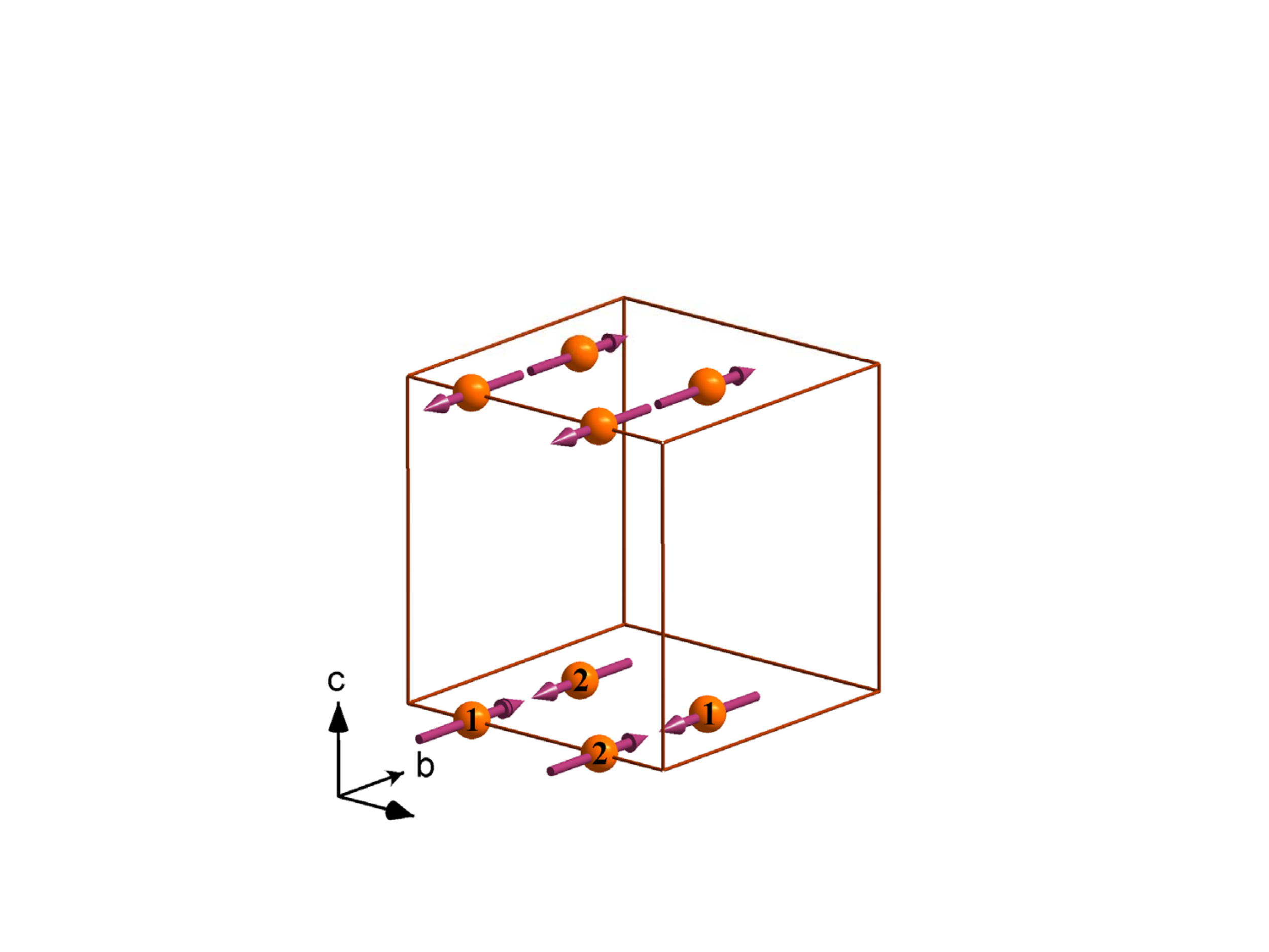}\\
\vspace{-6cm}
\flushleft (a)
\vspace{10cm}
\caption{(color online) Possible magnetic structures of FeSe$_{1-x}$ under pressure: (a) $m_y(K_{I})$-type and (b) $m_y(K_{II})$-type. $m_i$ are the the iron magnetic order parameters (see Eq. (\ref{eq_magnetic-orderparameter}).}
\label{fig_possible_magn_str}
\end{figure}

The calculations of the symmetry analysis and the magnitude and symmetry of the dipole fields of the Fe subsystem at the muon are more rigorously discussed in the Appendix. Application of pressure leads to an increase of the magnetic field at the muon stopping site as observed in the experiments (see Fig.~\ref{fig_musr_complete}) only for the $K_{I}$ and $K_{II}$ translation symmetries. For the $K_{0}$ and $K_{III}$ translation symmetries application of pressure would lead to a decrease of the magnetic field. This behavior can be explained as the result of a competition between a general constraint of the lattice constants and a simultaneous shifting of the muon positions further away from the Fe $ab$-plane. As a result from this feature and the above mentioned similarity to the $R$FeAsO family it may be concluded that only the $K_{I}$ and $K_{II}$ translation symmetries are possible symmetries of the magnetic structures for FeSe$_{1-x}$ under pressure. Comparing both possible magnetic structures $K_{I}$ and $K_{II}$ (shown in Fig.~\ref{fig_possible_magn_str}) with the experimental data presented in Fig.~\ref{fig_musr_complete} leads to magnetic fields along the $z$-coordinate of $B_z(K_{I})=354.6\cdot m_y(K_{I})$ and $B_z(K_{II})=334.3\cdot m_y(K_{II})$, respectively. Here $m_i$ are the iron magnetic order parameters (see Eq. (\ref{eq_magnetic-orderparameter}). This corresponds to a Fe magnetic moment $\mu\approx0.2\mu_B$ for both magnetic structures. 
However, the very modest shift of the muon position in the region $0.25-4$\,GPa calculated here, cannot explain the giant increase (four times) of the internal magnetic field $B_{\rm int}$ with an increase of the pressure from 1\,GPa to 2.4\,GPa. Therefore, all these changes are connected with a pressure induced increase of the iron magnetic moment. The right scale of Fig.~\ref{fig_musr_complete}c shows the estimated value of the magnetic  moment using dipole-dipole calculations for $4$\,GPa.

\section{Neutron Diffraction}

Neutron diffraction experiments were performed on the Cold Neutron Powder Diffractometer DMC at SINQ (PSI) at a pressure of $p=4.4(5)$\,GPa in a Paris-Edinburgh press\cite{Klotz_APL_2005} in order to investigate the proposed magnetic structures of FeSe$_{1-x}$ on polycristalline samples of 40\,mm$^3$ effective volume in the beam. The pressure was determined by the known pressure dependence of the $c$-axis of FeSe$_{1-x}$.\cite{Margadonna_PRB_09}
The experiments were performed at temperatures of 5\,K and 150\,K using neutrons with a wavelength of $\lambda=2.4575$\,\AA. The FULLPROF program was applied to analyze and to model the diffraction data.\cite{Rodriguez_PhysicaB_1993} 

\begin{figure}[t!]
\centering
\vspace{-0cm}
\includegraphics[width=\linewidth]{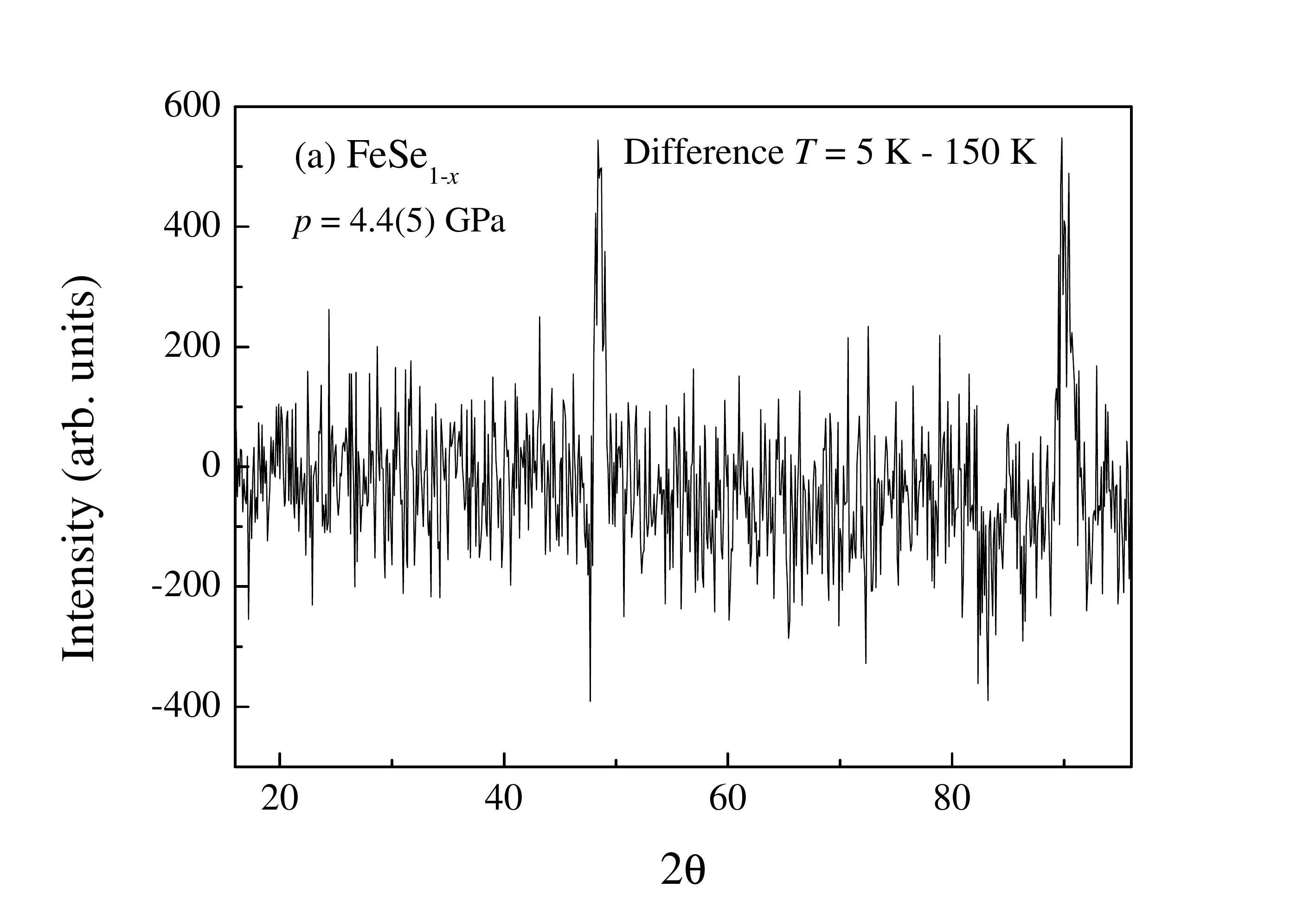}
\includegraphics[width=\linewidth]{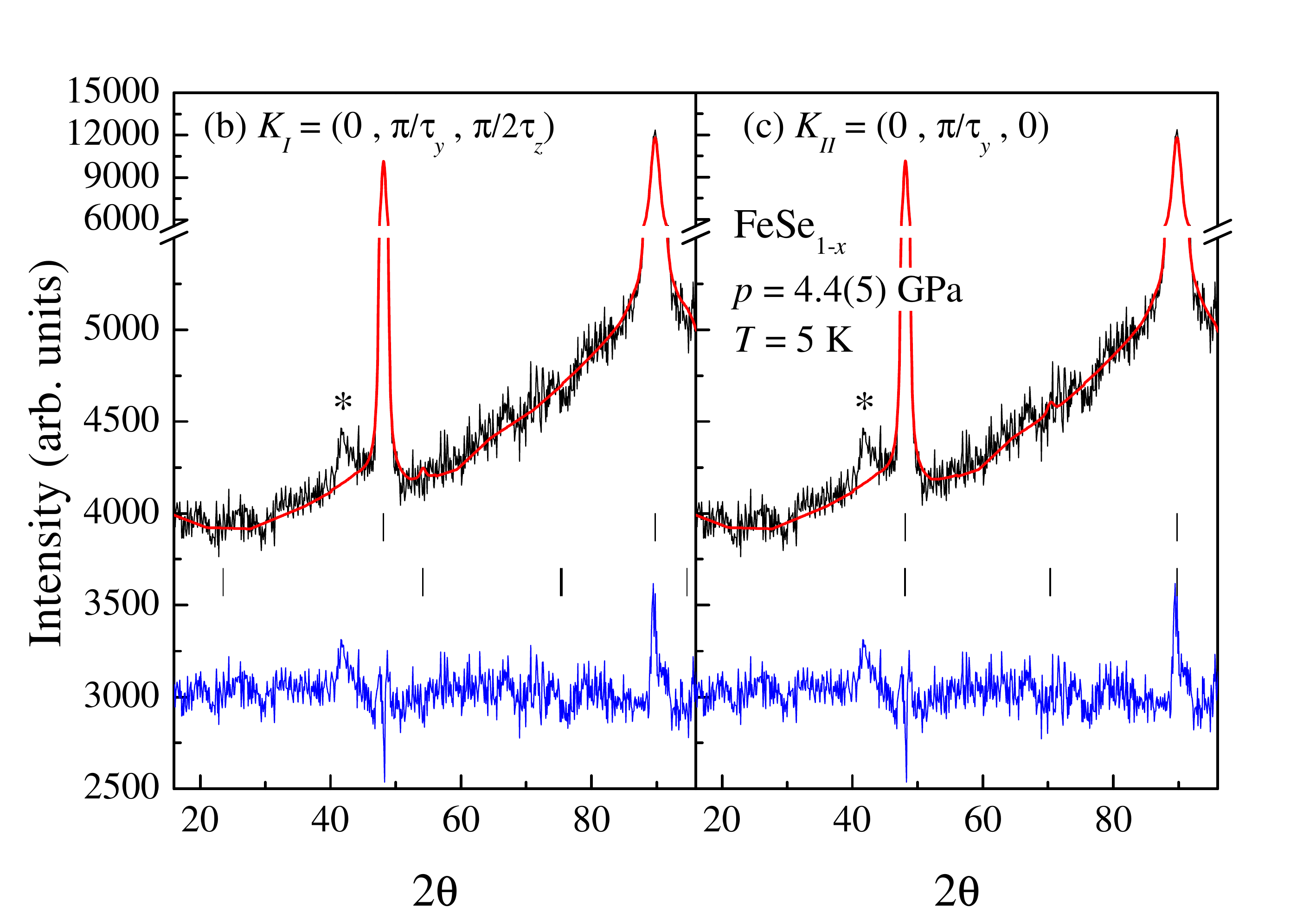}
\caption{(color online) (a) Difference of the neutron diffraction spectra of FeSe$_{1-x}$ taken at $T=5$\,K and 150\,K. Only noise is seen, except at the position of the nuclear peaks due to different Debye-Waller factors. Simulations of the magnetic structures (b) $m_y(K_{I})$-type and (c) $m_y(K_{II})$-type to the measured neutron diffraction spectra at $T=5$\,K for a moment of 0.5$\,\mu_{\rm B}$ per iron atom. The possible magnetic diffraction peaks are hidden in the background signal for all magnetic structures proposed. The peak in (b) and (c) indicated with {\large $\ast$} is a temperature independent feature of the pressure cell.}
\label{fig_neutron_simulation}
\end{figure}

The diffraction patterns measured at $T=5$\,K and $150$\,K were normalized to each other, and then subtracted from each other in order to obtain evidence of possible magnetic Bragg peak. However, no difference peak was observed, except at the positions of the nuclear peaks (see Fig.~\ref{fig_neutron_simulation}a). 
The different intensities of the nuclear peaks at the investigated temperatures result from the temperature dependent Debye-Waller factors. There are two possible explanations that no magnetic Bragg peaks were observed with neutrons in contrast to $\mu$SR which shows static magnetism: (i) the magnetic moment is too small, resulting in an intensity of the magnetic diffraction peak that is hidden below the background of the sample and the pressure cell, or (ii) the magnetic order is static, but no long range order occurs (muons are sensitive only over a few unit cells). 

However, because oscillations are seen in the $\mu$SR time spectra (see Fig.~\ref{fig_musr_raw}) the magnetic order is long range, thus leading to the conclusion that static magnetic order occurs below $T_{\rm N}$. The muon stopping site calculations have shown that the magnetic moment is quite small ($\approx0.2\mu_B$/Fe at $p=2.4$\,GPa). A linear exrtrapolation of the moment with pressure would lead to a moment of $\approx0.35\mu_B$ at $p=4.4$\,GPa. Therefore, we analyzed the neutron data using a theoretical model considering the two proposed magnetic structures $K_I$ and $K_{II}$. For both structures the magnetic peaks are hidden in the background. The simulated diffraction patterns for the strutures with the magnetic vector $K_I$ and $K_{II}$ are shown in Fig.~\ref{fig_neutron_simulation}b and c. The largest possible magnetic moment, that is not seen due to the high background of the pressure cell is estimated to $\approx0.5-0.7\,\mu_{\rm B}$ per iron atom (dependent on the magnetic structure). The simulations of the estimated structures are in agreement with the muon stopping site calculations that show a very low magnetic moment per Fe atom.  

\section{Phase Diagram}

\begin{figure}[t!]
\centering
\vspace{-0cm}
\includegraphics[width=1\linewidth]{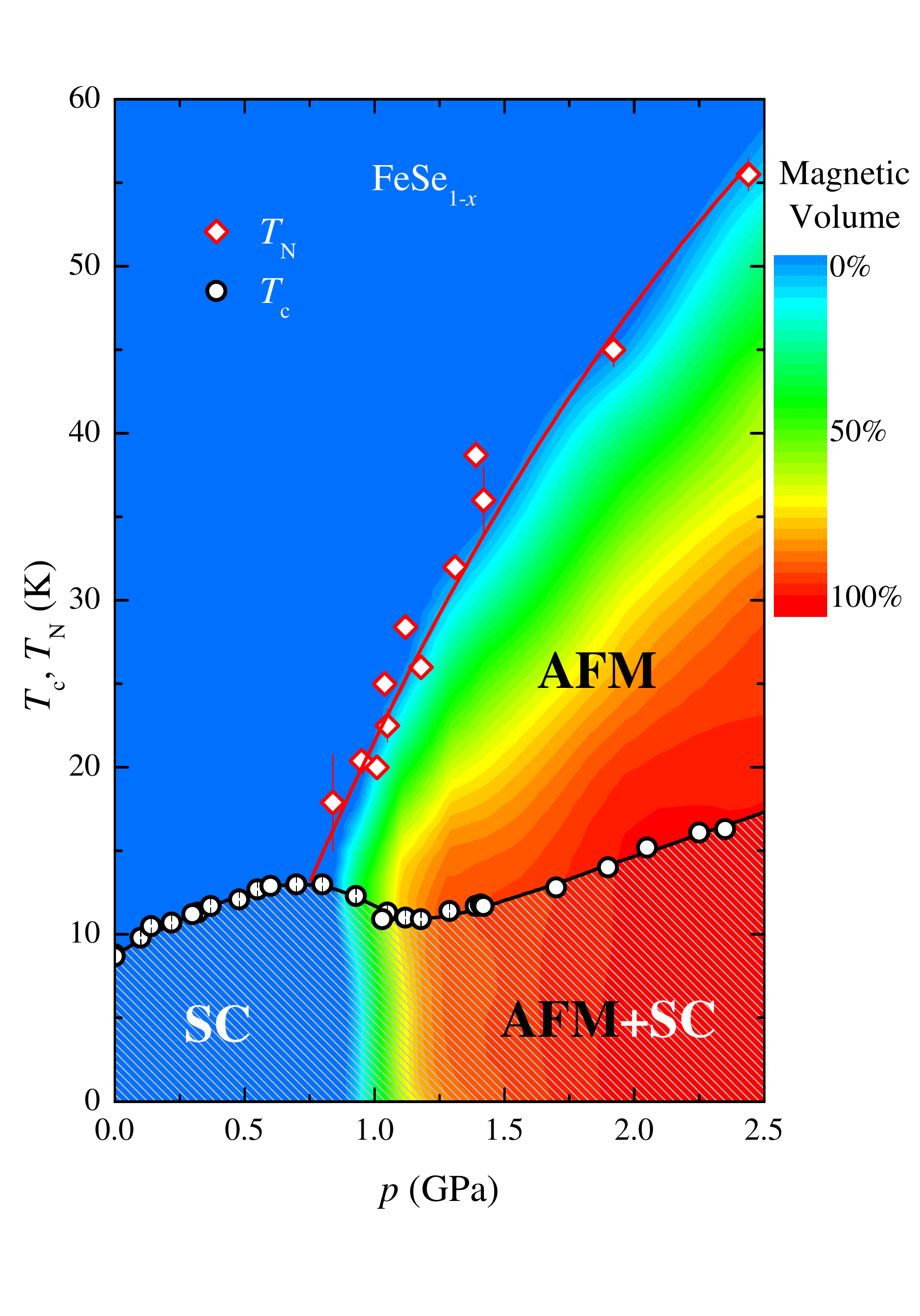}
\caption{(color online) Pressure dependence of the superconducting transition temperature $T_{\rm c}$, the magnetic ordering temperature $T_{\rm N}$, and the superconducting and magnetic volume fractions of FeSe$_{1-x}$. The superconducting volume is $100$\% for all pressures investigated, determined from ac susceptibility and muon spin rotation experiments of FeSe$_{1-x}$. The data obtained in this study are plotted together with the data from Refs.~\onlinecite{Bendele_PRL_10} and 
\onlinecite{Ichsanow_2010}. The $T_{\rm c}$ and $T_{\rm N}$ lines are guides to the eye and SC, M, and PM denote the superconducting, magnetic and nonmagnetic states of the samples, respectively.  }
\label{fig_phasediag}
\end{figure}

Figure~\ref{fig_phasediag} summarizes the results obtained in this study in a phase diagram. At low pressures below $p\leq 0.8$\,GPa the samples are superconducting only and show an increase of $T_{\rm c}$ from $\sim8$\,K at ambient pressure to $\sim13$\,K at $\simeq0.8$\,GPa. At higher pressures static magnetic order is established below $T_{\rm N}>T_{\rm c}$ that first competes and coexists with superconductivity, and at higher pressure ($p\gtrsim1.2$\,GPa) it only coexists with superconductivity. In the intermediate pressure range ($0.8\leq p \leq 1.2$\,PGa) the competition is evident from two observations: (i) as a function of pressure $T_{\rm c}$ is suppressed as soon as mangetic order appears, leading to the local maximum of $T_{\rm c}$ at $p\simeq 0.8$\,GPa. However, the superconducting volume fraction remains to be $100$\%. (ii) the magnetic order, that is established above $T_{\rm c}$ is parially (or even fully)\cite{Bendele_PRL_10} suppressed by the onset of superconductivity. This is seen by a decrease of the internal magnetic field $B_{\rm int}(0)$ and a decrease of the magnetic volume fraction when the samples enter the superconducting state (see Fig.~\ref{fig_musr_data} and Ref.~\onlinecite{Bendele_PRL_10}). For $p\gtrsim1.2$\,GPa magnetism is fully established, and both $T_{\rm N}$ and the magnetic moment increase with increaing pressure. Interestingly, the onset of magnetic order and the simultaneous rapid increase of the Fe magnetic moment coincide with a drastic change of the Se height above the Fe plane that starts also at $\sim 1$\,GPa.\cite{Margadonna_PRB_09,Mizuguchi_SST_10} 

The appearance of antiferromagnetic order has also been seen by NMR measurements.\cite{Imai_PRL_09} An increase of $1/TT_1$ close to $T_{\rm c}$ is observed at low pressures ($p=0$ and 0.7\,GPa) indicating antiferromagnetic modes of spin fluctuations that are strongly enhanced towards $T_{\rm c}$. This leads to the conclusion that FeSe$_{1-x}$ is in close proximity to a magnetic instability. 
At higher pressures (at 1.4\,GPa and 2.2\,GPa, \textit{i.e.}~where $\mu$SR observes static magnetic ordering) the $1/TT_1$ data reveal a broad hump significantly above $T_{\rm c}$. Furthermore, the integrated intensity of the NMR signal begins to decrease at about $34$\,K at 1.4\,GPa and at about $50$\,K at 2.2\,GPa, in excellent agreement with the $\mu$SR data. The disappearance of the NMR signal below a peak of $1/TT_1$ is a characteristic signal for a magnetic phase transition with a (nearly) static magnetic hyperfine field with a broad distribution.\cite{Imai_PRL_09}

Keeping in mind that the superconducting volume fraction is $\simeq100$\% for all pressures measured and that the magnetic volume fraction reaches $\simeq100$\% at $p\gtrsim1.2$\,GPa indicates that both ground states coexist in the whole sample volume. The data do not show any signature for macroscopic phase separation into superconducting and magnetic regions larger than a few nanometers, as observed \textit{e.g.}~in Ba$_{1-x}$K$_{x}$Fe$_2$As$_2$\cite{Park_PRL_09} or LaFeAsO$_{1-x}$F$_x$.\cite{Khasanov_arxiv_LaFeAs_11} No sublattice is present which could order magnetically, while the superconducting FeAs layers are not magnetically ordered, as \textit{e.g.}~observed in Ce1111 or Sm1111.\cite{Maeter_PRB_09,Khasanov_PRB_SmFeAs_08} These observations point rather to an atomic scale coexistence of the order parameters as it is seen \textit{e.g.}~in FeTe$_{1-x}$Se$_{x}$\cite{Khasanov_PRB_FeTeSe_09} or Ba(Fe$_{1-x}$Co$_{x}$)$_2$As$_2$.\cite{Laplace_PRB_09} Furthermore, it seems that the two ground states stabilize each other with pressure as $T_{\rm c}$, $T_{\rm N}$, and $B_{\rm int}(0)$ are increasing in parallel with increasing pressure. Comparing FeSe$_{1-x}$ with the newly discovered $R$Fe$_{2-x}$Se$_2$ (245) system in which superconductivity and magnetism coexist rises the question, whether magnetic order in FeSe$_{1-x}$ under pressure is of similar origin as the one in the 245 system.\cite{Shermadini_PRL_2011,Hu_arxiv_2011} In the latter system the superconducting transition temperatures reaches $T_{\rm c}\simeq32$\,K and superconductivity seems to coexists with magnetism occuring at $T_{\rm N}\approx 500$\,K with a rather large magnetic moment of $3\mu_B$ per Fe atom.\cite{Bao_arxiv_2011} 

Knowing that FeSe$_{1-x}$ is a two gap superconductor\cite{Khasanov_PRB_08,Khasanov_PRL_10} a possible scenario of an atomic scale coexistence of superconductivity and magnetism has recently been proposed by Vorontsov \textit{et al.}\cite{Vorontsov_PRB_2009,Vorontsov_PRB_2010,Vavilov_SST_2010} and Cvetkovic and Tesanovic.\cite{Cvetkovic_PRB_09} They proposed a region in which superconductivity and magnetic order can coexist. Here, the magnetic order can be commensurate only in a rather small parameter range where the Fermi surface nesting is not perfect. The bands are supposed to have an elliptical shape, and the chemical potential is supposed to shift.

\section{Conclusions}
The pressure dependence of the superconducting and magnetic properties of FeSe$_{1-x}$ were studied by means of ac and dc magnetization, as well as zero field $\mu$SR techniques. It is shown that independent on the preparation procedure the samples are bulk superconductors up to a  pressure of $p\simeq2.4$\,GPa. The superconducting transition temperature $T_{\rm c}$ increases with increasing pressure. However, the increase is non linear: $T_{\rm c}$ exhibits a local maximum at $0.8$\,GPa and a local minimum at $1.2$\,GPa. At pressures higher than $\simeq0.8$\,GPa static magnetic ordering occurs below the N\'eel temperature $T_{\rm N}>T>T_{\rm c}$. In an intermediate pressure range where $T_{\rm c}$ is decreasing ($0.8\leq p\leq1.2$\,GPa) the magnetic order is incommensurate and competes with superconductivity.\cite{Bendele_PRL_10} Only at $p\gtrsim1.2$\,GPa when magnetic order is fully established, the magnetic order is commensurate and magnetism occupies the full sample volume, coexisting with superconductivity on an atomic length scale. Muon stopping site calculations reveal only one stopping site of the muons along the Se - Se connection and a small pressure dependent magnetic moment with a value of $\sim0.2\mu_B$ at $p\sim2.4$\,GPa is found. A recent M\"ossbauer study reported no magnetic order in FeSe$_{1-x}$.\cite{Medvedev_Nat_09} However, the samples were prepared in a slightly different way. Following carefully the preparation procedure used in the M\"ossbauer study and investigating these samples by means of $\mu$SR, clear evidence of magnetic order in the system is observed,\cite{Ichsanow_2010} in contrast to the M\"ossbauer results.\cite{Medvedev_Nat_09} 

Different magnetic structures based on the muon stopping site calculations and a symmetry anlaysis are proposed and tested. The neutron diffraction measurements did not reveal any magnetic Bragg reflections because the magnetic moment seems to be too small. Thus, only speculations about the magnetic structure are possible. It is most probably very similar to the magnetic structure of the LaFeAsO family of Fe-based superconductors, since the FeSe$_{1-x}$ layers resemble the FeAs layers in the $R$1111 system. 

Both superconductivity and magnetism are stabilized by pressure. This is evident from the simultaneous increase of $T_{\rm c}$, $T_{\rm N}$, and $B_{\rm int}(0)$ and the related magnetic moment $\mu$ with increasing pressure. It remains to be seen whether this peculiar behavior influences or even helps to clarify the pairing mechanism in the Fe-based superconductors.

\section{Acknowledgment}
This work was supported by the Swiss National Science Foundation. Yu.~Pashkevich acknowledges partial support from the Swiss National Science Foundation (grant SNSF IZKOZ2\_134161). The work at Donetsk PhysTech has been supported under Ukrainian-Russian Grant No. 9-2010 and NASU Grant No.232. The experiments were partially performed at the Swiss Muon Source S$\mu$S and at the Swiss neutron spallation SINQ of the Paul Scherrer Institute PSI, Switzerland. Helpful discussions with V. Yu. Pomjakushin and S. Weyeneth are acknowledged.

\subsection{Magnitude and symmetry of dipole fields from Fe subsystems at the muon site}
\begin{center}
\begin{table*}[t]\centering\caption{Symmetry of iron magnetic order parameters ($m_i$ and $l_i$ that are a linear combination of the sublattice moments, see Eq.~(\ref{eq_magnetic-orderparameter})), and the corresponding magnetic fields at the muon sites ($M_i$ and $L_i$, see Eq.~(\ref{eq_magnetic-moment})) in FeSe$_{1-x}$ for the four possible propagation vectors of magnetic ordering $K_l$ ($l=0,\,I,\,II,\,III$). The enumeration of the irreproducable representations (IR) $\tau_i$ is given in accordance with the Kovalev notation.\cite{Kovalev_93} }\hspace{20mm}
\hfill{}
 \begin{tabular}{|c| c| c| c| c| c| c| c| c|} \toprule
   & \multicolumn{2}{c|}{$K_0=(0,0,0)$} & \multicolumn{2}{c|}{$K_{I}=(0,\pi/\tau_y,\pi/2\tau_z)$} & \multicolumn{2}{c|}{$K_{II}=(0,\pi/\tau_y,0)$} & \multicolumn{2}{c|}{$K_{III}=(0,0,\pi/2\tau_z)$}\\ \hline

   IR & \parbox{2cm}{Fe-order parameters} & \parbox{2cm}{fields at $\mu^+$ site} & \parbox{2cm}{Fe-order parameters} & \parbox{2cm}{fields at $\mu^+$ site} & \parbox{2cm}{Fe-order parameters} & \parbox{2cm}{fields at $\mu^+$ site} & \parbox{2cm}{Fe-order parameters} & \parbox{2cm}{fields at $\mu^+$ site} \\ \hline
  
   $\tau_1$& -- & -- & $m_x$ & -- & $m_x$ & -- & -- & -- \\ \hline
   $\tau_2$& -- & $L_z$ & $l_x$ & $M_z$ & $l_x$ & $L_z$ & -- & $M_z$ \\ \hline
   $\tau_3$& $m_x$ & $M_x$ & $--$ & $L_x$ & $--$ & $M_x$ & $m_x$ & $L_x$ \\ \hline
   $\tau_4$& $l_x$ & $L_y$ & $--$ & $M_y$ & $--$ & $L_y$& $m_x$ & $M_y$ \\ \hline
   $\tau_5$& $m_y$ & $M_y$ & $m_z$ & $L_y$ & $m_z$ & $M_y$& $l_x$ & $L_y$ \\ \hline
   $\tau_6$& $l_y$ & $L_x$ & $l_z$ & $M_x$ & $l_z$ & $L_x$& $m_y$ & $M_x$ \\ \hline
   $\tau_7$& $m_z$ & $M_z$ & $m_y$ & $L_z$ & $m_y$ & $M_z$& $l_y$ & $L_z$ \\ \hline
   $\tau_8$& $l_z$ & $--$ & $l_y$ & $--$ & $l_y$ & $--$& $l_z$ & $--$ \\ \toprule
  \end{tabular}\hfill{}
\label{tab_symmetry}
\end{table*}
\end{center}

The symmetry analysis was done assuming that the overall distribution of the magnetic fields in the magnetic unit cell has the same symmetry as the magnetic order parameter. In order to find the orientation of the magnetic field at the muon site, an artifacial magnetic moment is ascribed to this site. The corresponding set of magnetic degrees of freedom forms the magnetic representation for some positions (Wyckoff positions). The magnetic representation is transferred into an irreducible representation $\tau_i$ after making a standard decomposition. After that it is possible to analyze the possible symmetry of the mangetic moment (i.e. staggered magnetic fields) at the muon site.

The magnetic order parameters consist of Fourier components of respective magnetic propagation vectors $K_l$ of the $\alpha$ sublattice magnetic moments $m_i^{(\alpha)}(K_l)$ ($\alpha=1,2$):
\begin{align}
 \nonumber m_i(K_l)=&\frac{1}{2}\left(m_i^{(1)}(K_l)+m_i^{(2)}(K_l) \right); \\l_i(K_l)=&\frac{1}{2}\left(m_i^{(1)}(K_l)-m_i^{(2)}(K_l) \right); \hspace{10pt}l=0,I,II,III.
\label{eq_magnetic-orderparameter}
\end{align}
The nonzero components of respective magnetic moments at the muons sites have the form: 
\begin{align}
 \nonumber M_i(K_l)=&\frac{1}{2}\left(B_i^{(1)}(K_l)+B_i^{(2)}(K_l) \right); \\L_i(K_l)=&\frac{1}{2}\left(B_i^{(1)}(K_l)-B_i^{(2)}(K_l) \right); \hspace{10pt}l=0,I,II,III.
\label{eq_magnetic-moment}
\end{align}
Here $B_i^{(\alpha)}(K_l)$ is the $i$-cartesian component of a magnetic field at the muon site $\alpha$ $(\alpha=1,2)$ with $K_l$ type symmetry.

In Table~\ref{tab_symmetry} the result of the symmetry analysis is presented. Here, the enumeration of the irreproducible representations $\tau_i$ is given in accordance with the Kovalev notation.\cite{Kovalev_93} It shows the symmetry of the iron magnetic order parameter $m_i$ and $l_i$, and the corresponding magnetic fields at the muon sites $M_i$ and $L_i$ for the four magnetic propagation vectors $K_l$. Due to the high local symmetry of the muon sites some directions of the iron magnetic structure cannot create a magnetic field at the muon sites. Thus, the observation of $\mu$SR signals (oscillations in the $\mu$SR time spectra, see Fig.~\ref{fig_musr_raw}) at high pressures in FeSe$_{1-x}$ evidences that the magnetic structure has a certain direction and a certain arrangement of exchange interactions (\textit{i.e.}~different type of exchange order). 

The analysis of the magnitude and the symmetry of the dipole fields for the possible propagation vectors of magnetic ordering $K_l$ ($l=0,\,I,\,II,\,III$) at the muon site of the Fe subsystems in FeSe$_{1-x}$ leads to the results obtained in Eqs.~(\ref{eq_matrix1}) and (\ref{eq_matrix2}). There the magnetic fields are given in mT, and the basis functions ($m$ and $l$) in the units of $\mu_B$. 

\begin{widetext}
For $4$\,GPa the following results were obtained:
\begin{eqnarray}
\nonumber 
\begin{pmatrix} B_x(K_{I})\\B_y(K_{I})\\B_z(K_{I}) \end{pmatrix} & = &
\begin{pmatrix} 0&0&0 \\ 0&0&354.6 \\ 0&354.6&0 \end{pmatrix} \begin{pmatrix} m_x(K_{I})\\m_y(K_{I})\\m_z(K_{I})\end{pmatrix}+
\begin{pmatrix} 0&0&-351.1\\0&0&0\\-351.1&0&0\end{pmatrix} \begin{pmatrix} l_x(K_{I})\\l_y(K_{I})\\l_z(K_{I})\end{pmatrix}\\
\nonumber
\begin{pmatrix} B_x(K_{II})\\B_y(K_{II})\\B_z(K_{II}) \end{pmatrix} & = &
\begin{pmatrix} 0&0&0\\0&0&-334.3\\0&-334.3&0\end{pmatrix} \begin{pmatrix} m_x(K_{II})\\m_y(K_{II})\\m_z(K_{II})\end{pmatrix}+
\begin{pmatrix} 0&0&331.3\\0&0&0\\331.3&0&0\end{pmatrix} \begin{pmatrix} l_x(K_{II})\\l_y(K_{II})\\l_z(K_{II})\end{pmatrix}\\
\nonumber
\begin{pmatrix} B_x(K_{0})\\B_y(K_{0})\\B_z(K_{0}) \end{pmatrix} & = &
\begin{pmatrix} 106.2&0&0\\0&111.0&0\\0&0&439.7\end{pmatrix} \begin{pmatrix} m_x(K_{0})\\m_y(K_{0})\\m_z(K_{0})\end{pmatrix}+
\begin{pmatrix} 0&-479.9&0\\-479.9&0&0\\0&0&0\end{pmatrix} \begin{pmatrix} l_x(K_{0})\\l_y(K_{0})\\l_z(K_{0})\end{pmatrix}\\
\begin{pmatrix} B_x(K_{III})\\B_y(K_{III})\\B_z(K_{III}) \end{pmatrix} & = & 
\begin{pmatrix} -217.0&0&0\\0&-222.7&0\\0&0&439.7\end{pmatrix} \begin{pmatrix} m_x(K_{III})\\m_y(K_{III})\\m_z(K_{III})\end{pmatrix}+
\begin{pmatrix} 0&476.1&0\\476.1&0&0\\0&0&0\end{pmatrix} \begin{pmatrix} l_x(K_{III})\\l_y(K_{III})\\l_z(K_{III})\end{pmatrix}
\label{eq_matrix1}
\end{eqnarray}
For $9$\,GPa the following results were obtained:
\begin{eqnarray}
\nonumber\begin{pmatrix} B_x(K_{I})\\B_y(K_{I})\\B_z(K_{I}) \end{pmatrix} &= &
\begin{pmatrix} 0&0&0\\0&0&374.7\\0&374.7&0\end{pmatrix} \begin{pmatrix} m_x(K_{I})\\m_y(K_{I})\\m_z(K_{I})\end{pmatrix}+
\begin{pmatrix} 0&0&-371.5\\0&0&0\\-371.5&0&0\end{pmatrix} \begin{pmatrix} l_x(K_{I})\\l_y(K_{I})\\l_z(K_{I})\end{pmatrix}\\
\nonumber\begin{pmatrix} B_x(K_{II}\\B_y(K_{II})\\B_z(K_{II}) \end{pmatrix} &=& 
\begin{pmatrix} 0&0&0\\0&0&-348.1\\0&-348.1&0\end{pmatrix} \begin{pmatrix} m_x(K_{II})\\m_y(K_{II})\\m_z(K_{II})\end{pmatrix}+
\begin{pmatrix} 0&0&345.4\\0&0&0\\345.4&0&0\end{pmatrix} \begin{pmatrix} l_x(K_{II})\\l_y(K_{II})\\l_z(K_{II})\end{pmatrix}\\
\nonumber\begin{pmatrix} B_x(K_{0})\\B_y(K_{0})\\B_z(K_{0}) \end{pmatrix} &=& 
\begin{pmatrix} 82.1&0&0\\0&86.6&0\\0&0&-168.7\end{pmatrix} \begin{pmatrix} m_x(K_{0})\\m_y(K_{0})\\m_z(K_{0})\end{pmatrix}+
\begin{pmatrix} 0&-456.2&0\\-456.2&0&0\\0&0&0\end{pmatrix} \begin{pmatrix} l_x(K_{0})\\l_y(K_{0})\\l_z(K_{0})\end{pmatrix}\\
\begin{pmatrix} B_x(K_{III})\\B_y(K_{III})\\B_z(K_{III}) \end{pmatrix} &= &
\begin{pmatrix} -200.6&0&0\\0&-205.2&0\\0&0&405.8\end{pmatrix} \begin{pmatrix} m_x(K_{III})\\m_y(K_{III})\\m_z(K_{III})\end{pmatrix}+
\begin{pmatrix} 0&450.8&0\\450.8&0&0\\0&0&0\end{pmatrix} \begin{pmatrix} l_x(K_{III})\\l_y(K_{III})\\l_z(K_{III})\end{pmatrix}
\label{eq_matrix2}
\end{eqnarray}
\end{widetext}

\end{document}